\documentclass[pra,aps,superscriptaddress,balancelastpage,onecolumn]{revtex4}
\usepackage{amsmath}
\usepackage{graphicx}
\usepackage{amsfonts}
\usepackage{amssymb}
\usepackage{soul,xcolor}
\usepackage{lmodern}
\usepackage[utf8]{inputenc}
\usepackage[T1]{fontenc}
\providecommand{\U}[1]{\protect\rule{.1in}{.1in}}
\providecommand{\U}[1]{\protect\rule{.1in}{.1in}}

\begin{document}
\title{Light drag in an Optomechanical system}
\author{Hazrat Ali}
%\email{yamanuom@gmail.com}
\affiliation{Department Physics, Abbottabad University of Science and Technology, Havellian, 22500, KPK, Pakistan}
\author{Nadia Boutabba}
\affiliation{Fatima College of Health Sciences, Institute of Applied Technology, Abu Dhabi, UAE}
\affiliation{National Institute of Engineering of Carthage, University of Carthage, Tunis, Tunisia}
\author{Amjad Sohail\footnote{Corresponding author}}
\email{amjadsohail@gcuf.edu.pk}
\affiliation{Department of Physics, Government College University, Allama Iqbal Road, Faisalabad 38000, Pakistan}
\affiliation{Instituto de Física Gleb Wataghin, Universidade Estadual de Campinas, Campinas, SP, Brazil}
\begin{abstract}
Light dragging refers to the change in the path of light passing through a moving medium. This effect enables accurate detection of very slow speeds of light, which have prominent applications in state transfer, quantum gate operations, and quantum memory implementations. Here, to the best of our knowledge, we demonstrate the existence of the light-dragging effect in an optomechanical system (OMS) for the first time. The origin of this key factor arises from the nonlinear effects linked to optomechanical-induced transparency (OMIT). Hence, we observe prominent effects in the group and refractive indices profile spectra related to optomechanical parameters such as the decay rate of the cavity field, the mirror's damping momentum rate, and mechanical frequency. We find out that lateral light drag depends on the detuning by altering the amplitude and direction of the translational velocity. This allowed us to change the light's propagation through the optomechanical cavity from superluminal to subluminal and vice versa by modifying the probe's detuning. The ability to manipulate and control the light drag through an optomechanical system might be useful in designing novel optical devices and systems with enhanced performance.
\end{abstract}
\maketitle
%\pacs{}
\section{Introduction \label{secF}}
Following Einstein's theory of special relativity, the speed of light in a vacuum remains constant regardless of the reference frame. Nonetheless, the theoretical work of Fresnel, initially reported in 1818 \cite{fresnel1818influence}, demonstrated dragging in the light path which occurs when a light beam passes through a moving medium. This dragging can be either normal optical drag, which happens in the same direction as the motion of the medium, or anomalous optical drag, which happens in the opposite direction \cite{banerjee2022anomalous,qin2020fast,ullah2023coherent}. Fresnel observed that a light beam traveling at a transverse velocity of $v$ across the moving dielectric medium experiences a lateral displacement $\Delta x=(n_{g}-(n_{r})^{-1})(vL/c)$, where $n_{r}$($n_{g}$) is the phase (group) refractive index, $v$ ($c$) is the speed of moving medium (light), and $L$ is the moving length of the medium. Decades later, Fizeau empirically confirmed the changes in the speed of light in an experiment based on an interferometer and a tube filled with flowing water. Indeed, the findings indicated that the velocity of light through moving water is affected by its flow speed \cite{fizeau1991hypotheses}. Besides, the dispersion of light would influence the light-dragging effect as observed experimentally by Zeeman \cite{zeeman1919propagation,kox1993pieter,safari2016light}.

All of this research played a crucial role in advancing the understanding of Einstein's theory of special relativity. Nonetheless, in the case of low dispersion, this light-dragging is almost negligible; thus, for any observable results, the host media must have either a large velocity or a long traveling path to obtain a strong drag effect. Indeed, using large dispersion, light drag enhancement was achieved in addition to a large rotation sensing of 9 degrees per hour,
which is enough for metrology and motion-sensing
application \cite{qin2020fast}. Kuan et al. provided another relevant example by developing a light-drag velocimeter based on atomic systems. Their device has a sensitivity of two orders of magnitude greater than the velocity width of the atomic medium. In addition, a few theoretical proposals have recently been suggested in which the light-dragging effect is discussed in detail using the atomic medium \cite{ali2023light,boutabba2023light}.
This enhancement was achieved by exploiting the large dispersion of the electromagnetically induced transparency (EIT) medium, significantly increasing sensitivity compared to the Doppler width of the atomic ensemble \cite{kuan2016large}. This is because the optical drag effect is strongly related to the medium's dispersion characteristics and is therefore affected by several variables, such as resonance effects, absorption, and group and refractive indices. Processes like coherent population oscillation \cite{o2022random} or EIT  \cite{fresnel1818influence,radeonychev2020observation},  produce narrow resonances and consequently increase group indices and dispersion, which is necessary for the observation of slow light \cite{scully2003playing,sete2012controllable}. Thus, by employing EIT, the optical drag effect is further enhanced, approximately $10^5$ times greater than that of low dispersive media such as glass or water, allowing for a significant reduction in system sizes down to the millimeter range \cite{kuan2016large,banerjee2022anomalous}. However, practical applications are limited by the stringent experimental requirements for atomic systems, such as temperature regulation and vacuum conditions.

On the other hand, optomechanically induced transparency (OMIT) in optomechanical systems \cite{agarwal2010electromagnetically,huang2011electromagnetically}, which is analogous to EIT \cite{yan2020optomechanically}, holds significant promise for applications in quantum technologies, including quantum memories \cite{xiong2018fundamentals,kristensen2024long,weis2010optomechanically} and quantum entanglement repeaters \cite{yang2019optomechanically,koppenhofer2023single}.
OMIT has significant applications in the emergence of slow/fast light \cite{kocharovskaya2001stopping,turukhin2001observation, sohail2023rotational}, entanglement \cite{sohail2023enhanced23,ahmed2017optomechanical,ahmed2019effects}, the accurate determination of numerous physical quantities such as mass sensors \cite{he2015sensitivity}, and light storage \cite{lobino2009memory}.
The concept of OMIT is linked to the slowing of light \cite{yan2021optomechanically2}, which is key to the storage of quantum optical information and long-term memory processing  \cite{safavi2011electromagnetically,liu2019nonreciprocal,li2016transparency}. Within the transparency window, the group velocity of light is reduced.
Recently, Liu \textit{et. al.} demonstrated the diffraction grating in OMS and showed that it is easy to obtain the first, second, and third-order diffraction gratings using the light-mirror interaction strength \cite{WLM}.
A variety of mechanisms are involved \cite{akram2015tunable,jiang2017fano}; for instance, the detuning between the mechanical resonance and the optical field, as well as the interaction between the light and the moving mirror, create a narrow bandwidth where the light's group velocity is reduced, allowing it to be slowed significantly within the cavity. Herein, the moving resonator/membrane can act as a moving medium where the system might experience light drag.
Nonetheless, light drag has not yet been explored in a mechanical resonator coupled with an optical cavity, although it might offer a way to slow light. In this context, our research investigates the light drag in an optomechanical system wherein a mechanical resonator, with frequency $\omega_{m}$, is coupled to an optical cavity with frequency $\omega_{c}$. Our results provide a framework for further investigations to enhance the engineering of light-drag effects based on OMIT. Specifically, we demonstrate the quantum control of both group and refractive indices through optomechanical factors such as the momentum damping rate, the cavity decay rate (inversely related to cavity length), and the frequency of the mirror's oscillation.
Hence, we believe that the dragging of light in an optomechanical system can dynamically alter the refractive and group index, enabling tunable photonic devices and enhancing the sensitivity of optomechanical sensors by modulating the system's response to external perturbations.

The paper is organized as follows: in Section \ref{sec2}, the system's Hamiltonian and dynamical equations of the system are presented employing the standard quantum Langevin approach.
Next, we derive the light drag. The analysis is reported in section \ref{sec3}.
Finally, we present the concluding remarks in section \ref{sec4}.
\begin{figure}[b!]
\centering\includegraphics
%[width=0.5\linewidth]
[width=0.8\columnwidth,height=2.4in]{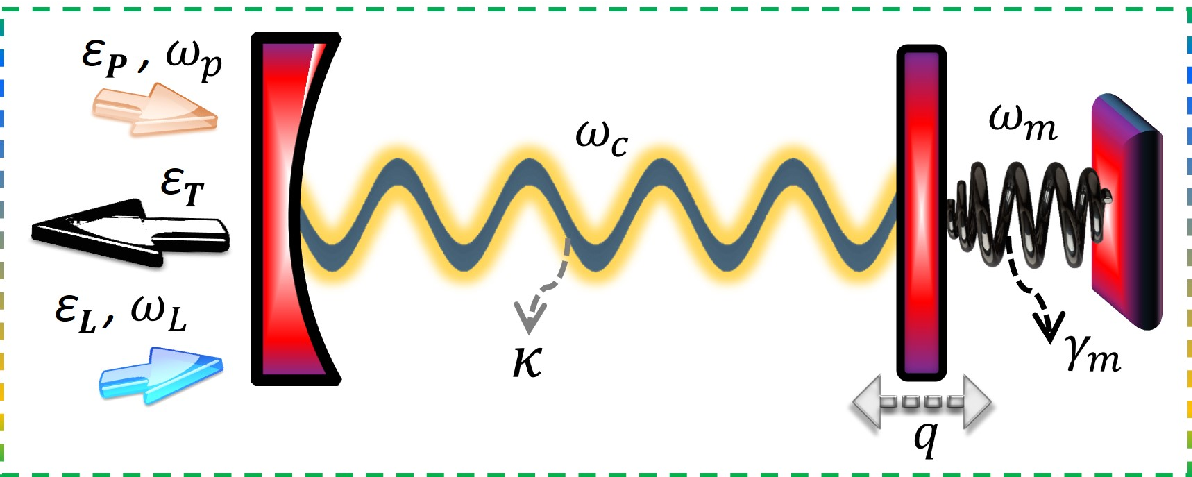}
\caption{Schematic diagram of a standard optomechanical system.}
\label{fig:model}
\end{figure}
\section{The Mechanical Resonator Coupling with the Optical Cavity\label{sec2}}

\label{sec2} We consider an optomechanical system, wherein a mechanical resonator, with frequency $\omega_{m}$, is coupled to an optical cavity with frequency $\omega_{c}$, as schematically shown in Fig. 1. The annihilation (creation) operator of the cavity field is denoted by $c$ $(c^{\dagger })$, satisfying the commutation relation $[c,c^{\dagger }]=1$, while $p$ and $q$ represents the momentum and position of the mechanical resonator of mass $m$. In an optomechanical cavity, one mirror can move freely as a result of the radiation pressure from light.
The detuning of the optical cavity is proportional to the displacement of a mechanical degree of freedom (e.g., mirror displacement or waveguide elongation, $q$).
Owing to the radiation pressure force that the photons within the cavity exert, the mechanical resonator experiences slight oscillations $q$. The cavity resonance frequency is then altered by the mechanical displacement. Therefore, we can expand the cavity frequency to the first order as $\omega_{c}(q)=\omega_{c}(1-\frac{q}{L})$. Thus, the interaction between the mechanical resonator and the cavity can be described by $-\hbar g_{m}c^{\dagger }c x$ with the optomechanical coupling constant as $g_m=\frac{\omega_{c}}{L}$, where $L$ is the cavity length. In the rotating frame at the coupling frequency $\omega _{L}$, the system's Hamiltonian can be expressed as:
\begin{widetext}
\begin{eqnarray}
\hat{H} &=&\hbar \Delta _{0}c^{\dagger }c+(\frac{p^{2}}{2m}+\frac{1}{2}%
m\omega _{m}^{2}q^{2})-\hbar g_{m}c^{\dagger }c x +i\hbar \epsilon _{L}\left( c^{\dagger }-c\right) +i\hbar \left(
c^{\dagger }\varepsilon _{p}e^{-i\delta t}-c\varepsilon _{p}^{\ast }e^{i\delta
t}\right), \label{HM}
\end{eqnarray}
\end{widetext}
where $\Delta _{0}=\omega _{c}-\omega _{L}$ is the detuning between the cavity's resonance frequency $\omega _{c}$ and the driving laser frequency $\omega _{L}$, and $\delta =\omega _{p}-\omega _{L}$, with $\omega _{c}, \omega _{p}$ being the resonance frequency of the optical cavity and the frequency of the probe field respectively.
In Eq. (\ref{HM}), the first term represents the bare Hamiltonian of the cavity field, and the second describes the bare Hamiltonian of the mechanical resonator of mass. The third term denotes the optomechanical coupling between the cavity mode and the mechanical resonator. The fourth gives the interaction between the driving laser and the optical cavity where
$\varepsilon_{L}$ is the amplitude of the driving laser field at frequency $\omega _{l}$. The last term describes the interaction of the probe field with the cavity and,
$\varepsilon_{p}$ is the amplitude of the probe field at frequency $\omega_{p}$, $\delta =\omega_{p}-\omega_{L}$ is the detuning between the probe and driving laser frequencies.
\subsection{Dynamics of the cavity optomechanical system}
In the mean-field approximation, the equations for mean values are given by:
\begin{eqnarray}
\left\langle \dot{q}\right\rangle &=&\frac{\left\langle p\right\rangle }{m}, \nonumber \\
\left\langle \dot{p}\right\rangle &=& -m\omega _{m}^{2}\left\langle q\right\rangle +\chi _{\circ }\left\langle c^{\dagger }\right\rangle \left\langle c\right\rangle -\gamma_{m} \left\langle p\right\rangle, \nonumber \\
\left\langle \dot{c}\right\rangle  &=&-\left[\kappa +i\left( \Delta _{0}-g_{m }q \right) \right] \left\langle c\right\rangle
+\varepsilon _{c}+\varepsilon _{p}e^{-i\delta t}. \label{OT}
\end{eqnarray}
Since Eq. (\ref{OT}) is inherently nonlinear, employing the standard linearisation technique would typically remove some of these nonlinear effects. However, given that the coupling field is significantly stronger than the probing field, we apply a perturbation method to solve Eq. (\ref{OT}) and obtain its steady-state solutions up to the first order in $\varepsilon_{p}$. i.e., $\left\langle f\right\rangle =f_{0}+f_{+}\varepsilon _{p}e^{-i\delta t}+f_{-}\varepsilon_{p}^{\ast }e^{i\delta t}$ $(f=q,p,c)$. Owing to the fact that the cavity and resonator have significantly strong coupling at the resonance frequency, i.e.,  here we take $\Delta \sim\delta \sim\omega _{m}$ and assume
 $x= \delta -\omega _{m}$. After some analytical calculations  (see the comprehensive analytical calculations in Appendix A), we arrive at the result of $\varepsilon _{T}$:
\begin{equation}
\varepsilon_{T}=\frac{2\kappa }{\kappa -ix+\frac{\beta }{\frac{\gamma_{m} }{2}-ix+\mathcal{N}}}, \label{ET}
\end{equation}
where%
\begin{equation}
\beta =\frac{g_{m}^{2}\varepsilon_{\circ }^{2}}{\Lambda(\kappa ^{2}+\omega_{m}^{2})}, \ \ \ \ \ (\Lambda=2m\omega_{m}/\hbar)
\end{equation}
and
\begin{equation}
\mathcal{N}=-\frac{\beta }{\kappa -2i\omega _{m}}.
\end{equation}
If we use the standard linearization approach to solve Eq. (2), the crucial term $\mathcal{N}$ will not appear in the subfraction of Eq. (\ref{ET}). Therefore, the origin of this key factor arises from the
nonlinear effects. Furthermore, the condition for ideal OMIT phenomena in a standard optomechanical system is fully explained in \cite{yan2020optomechanically}.
Furthermore, the condition for the pole's position in the subfraction can be determined from Eq. (3) by setting $\frac{\gamma_{m}}{2}-ix+\mathcal{N}=0$, we arrived at the two conditions:
\begin{eqnarray}
x=x_{0}&=&-\frac{\omega _{m}\gamma_{m}}{2\kappa},\\
\beta=\beta_{0}&=&\frac{\gamma(4\omega_{m}^{2}+\kappa^{2})}{2\kappa}.
\end{eqnarray}
The cavity's input-output relationship can be expressed as:
\begin{equation}
\varepsilon _{out}(t)+\varepsilon _{p}e^{-i\delta t}+\varepsilon _{c}=2\kappa c, \label{Eo}
\end{equation}
where
\begin{equation}
\varepsilon _{out}(t)=\varepsilon _{out}^{0}+\varepsilon _{out}^{+}\varepsilon _{p}e^{-i\delta t}+\varepsilon _{out}^{-}\varepsilon _{p}^{\ast}e^{i\delta t}.\label{Eo2}
\end{equation}
Solving Eq. (\ref{Eo}) and (\ref{Eo2}), one obtain
\begin{equation}
\varepsilon _{out}^{+}=\frac{2\kappa c_{+}-\varepsilon _{p}}{\varepsilon _{p}}.\label{Eo3}
\end{equation}
To make things easier, we define output field at the probe frequency
\begin{equation}
\varepsilon _{T}=\varepsilon _{out}^{+}+1=\frac{2\kappa c_{+}}{\varepsilon _{p}}.\label{Eo3}
\end{equation}
Here, $\varepsilon _{T}$ is a complex quantity that describes the quadrature of the field.
One can define the quadrature as $\chi=\chi_{r}+i\chi_{r}$  and can be obtained at the probe frequency, based on the method  \cite{baraillon2020linear}.
The real and imaginary terms display the absorption and dispersion spectrum of $\varepsilon _{T}$, respectively.
\subsection{Light drag effect in optomechanical system}
The novel idea is to discuss the light-dragging effects of OMS. The refractive and group indexes are the two major elements to discuss the light-dragging effect.
Since the output field is related to the optical susceptibility as $\chi=\varepsilon _{T}=\varepsilon _{out}^{+}+1$, the refractive index of the output field at the probe frequency can be defined as $n_r=1+2\pi\chi$. The group index at the probe field in OMS is given by
\begin{eqnarray}
n_{g}
&=&n_{r}+2\pi\omega \frac{\delta \chi}{\delta x},\nonumber \\
&=& 1+2\pi\chi+2\pi\omega \frac{\delta \chi}{\delta x}.
\end{eqnarray}
Here, the output field contains the real and imaginary parts, thus the refractive and group indexes of the optomechanical system have real and imaginary parts. The real part of the refractive index is related to the absorption, while the imaginary part describes the phase dispersion of light. The group velocity, delay, advancement, and attenuation can be obtained from the group index of the system.  From another side, the lateral light drag in the optomechanical system can be written as
\begin{equation}
\Delta_{x}=(n_{g}-\frac{1}{n_{r}})\frac{v l}{c}. \label{LLD}
\end{equation}
The parameters $v$, $c$, and $l$ are translation velocity, speed of light in
vacuum, and length of the medium, respectively. Moreover, it can be seen from Eq. (\ref{LLD})  that lateral light drag depends on both
the group refractive index $n_g$ and the phase refractive index $n_r$.
\begin{figure}[tbp]
\centering
  %\subfloat[The imaginary of refractive index versus probe detuning]
  \includegraphics[width=1\columnwidth,height=5in]{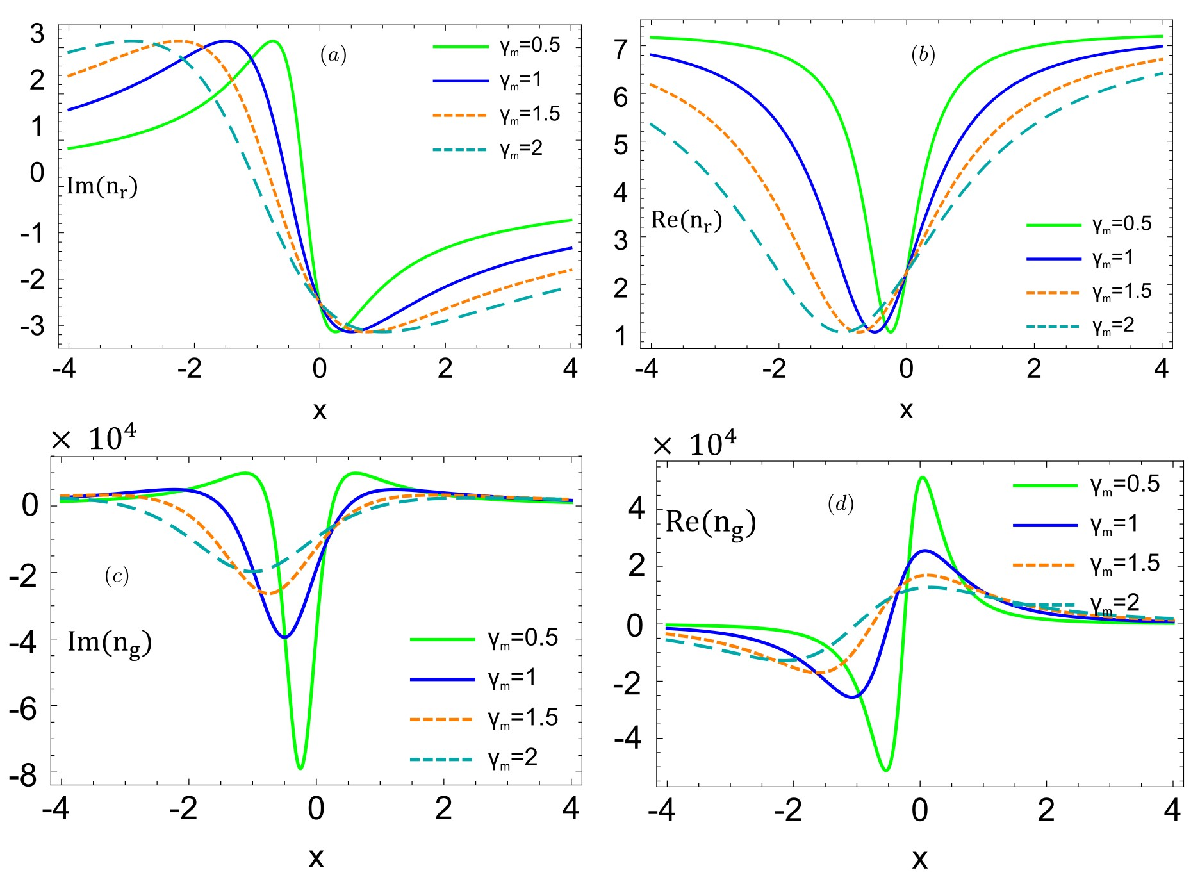}
%    \put(-110,150){($ a $)} \quad
% % \subfloat[The Real of refractive index ]
% { \includegraphics[width=.46\textwidth]{Fig2b.eps}}
% \put(-110,150){($ b $)} \quad \\
%  %\subfloat[The real part of group index index ]
%  { \includegraphics[width=.47\textwidth]{Fig2c.eps}}
%  \put(-200,120){($ c $)} \quad
%  %\subfloat[The imaginary part of the rgroup index ]
%  { \includegraphics[width=.47\textwidth]{Fig2d.eps}}
%   \put(-150,150){($ d $)} \quad
 \caption{(a)(c) The imaginary and (b)(d) the real part of the (a)(b) refractive index and (c)(d) the group index for different value of mechanical damping rate $\gamma_{m}$ are plotted against $x$. We take $\kappa=\omega_{m}$ \cite{{yan2020optomechanically}}.}
 \label{fig:2}
\end{figure}
\begin{figure}[b!]
\centering
  %\subfloat[The light drag versus y]
\includegraphics[width=1\columnwidth,height=2.7in]{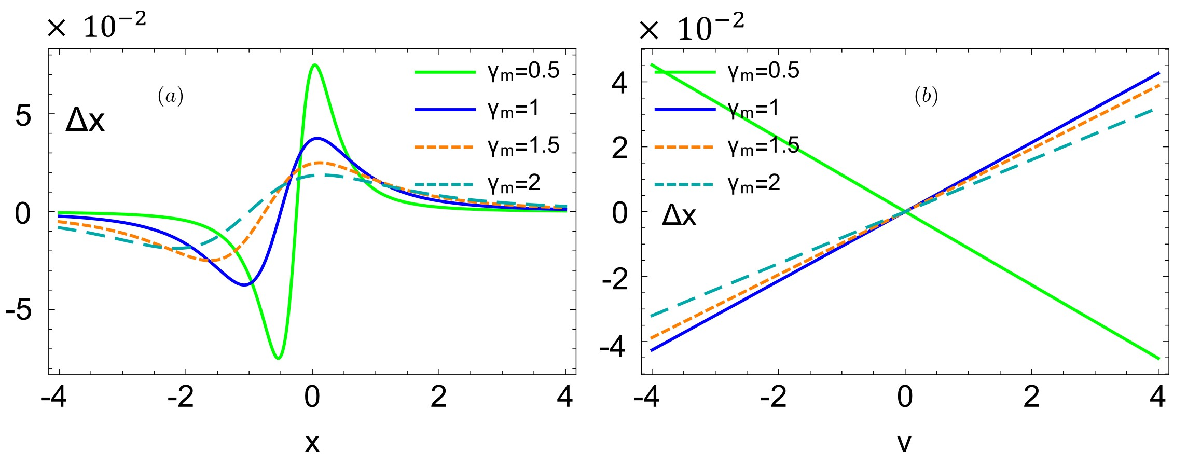}
 % \put(-180,150){($ a $)} \quad
  %\subfloat[The light darg versus v ]
%  {\includegraphics[width=.48\textwidth]{Fig3b.eps}}\put(-110,150){($ b $)} \quad
 \caption{The light drag in an optomechanical system as a function of (a) $x$ and (b) $v$  for different values of mechanical damping rate $\gamma_{m}$. The rest of the parameters are the same as in Fig. 2.}\label{fig:6}
\end{figure}

\section{Results and Discussion}\label{sec3}
In this section, we analyze the refractive index and the group index for various optomechanical factors, i.e., the coupling strength, the frequency of the mechanical resonator, the damping of the mechanical resonator, and the decay rate of the cavity. These factors influence various mechanisms of light and mechanical motion interactions, the lifetime of photons inside the cavity, and the rate at which the energy dissipates in the system. Hence, we report the light drag and examine the dispersion-absorption spectra. From Eq. (3), one can note that we can investigate the system by only three parameters, i.e., $\kappa$, $\omega_{m}$ and $\gamma_{m}$. In our numerical calculations, we have taken parameters $\kappa=\omega_{m}=10^{4}\gamma_{m}$ \cite{yan2020optomechanically}. It is also worth mentioning that the dip of the real part of Re($n_r$) and Im($n_g$) can be easily determined by Eq. (6). Moreover, the ideal dip of both Re($n_r$) and Im($n_g$) will appear around $x=-\frac{\gamma_{m}}{2}$, when $\kappa=\omega_{m}$, [see Eq. (6)].

In Fig. 2, we present the refractive index and group index as functions of probe detuning for different momentum damping rates, while keeping the cavity loss and mechanical frequency constant. The dispersion spectrum is illustrated by the imaginary part of the refractive index, whereas the real part of the refractive index represents the absorption spectrum.
In general, the damping momentum rate is related to the mechanical oscillation of the mirror, thereby affecting the group index (given by the velocity at which the envelope of a light pulse propagates) and the dispersion of the system.
Additionally, the mirror motion can induce a frequency shift in the cavity modes (through a Doppler-like effect). This frequency shift alters the group index since it is related to the derivative of the refractive index with respect to frequency. In Fig. 2, we see the refractive and group indices shift due to the high damping rate. Indeed, the slope of the anomalous dispersion (negative slope) curve is sharp around the resonance condition for a low momentum damping rate, i.e., $\gamma_{m}=0.5$, see the green curve of Fig. 2(a). For higher momentum damping rates, the anomalous dispersion changes ( see the blue, red, and cyan curves of Fig. 2(a), and the dispersion attenuates far from the resonance. In Fig. 2(b), the absorption spectrum is given by the real refractive index. We notice low absorption near the resonance for a small damping rate reflecting high transmission through the optomechanical cavity.  The low absorption dip is shifted towards negative probe detuning regions for high $\gamma_{m}$, as depicted by the green, blue, red, and cyan curves of Figure 2 (b).
Hence, we can suggest to shift the transmission by manipulating the detuning. For instance, for $\gamma_{m}=2$ , x=1, the absorption is about $57 \% $. By applying a phase shift $e^{i\frac{\pi}{2}}$ in the detuning, we can transmit the probe light to 90 $\%$ at x=-1.
\begin{figure}[tbp]
\centering
  %\subfloat[The imaginary of refractive index versus probe detuning]
  \includegraphics[width=1\columnwidth,height=5in]{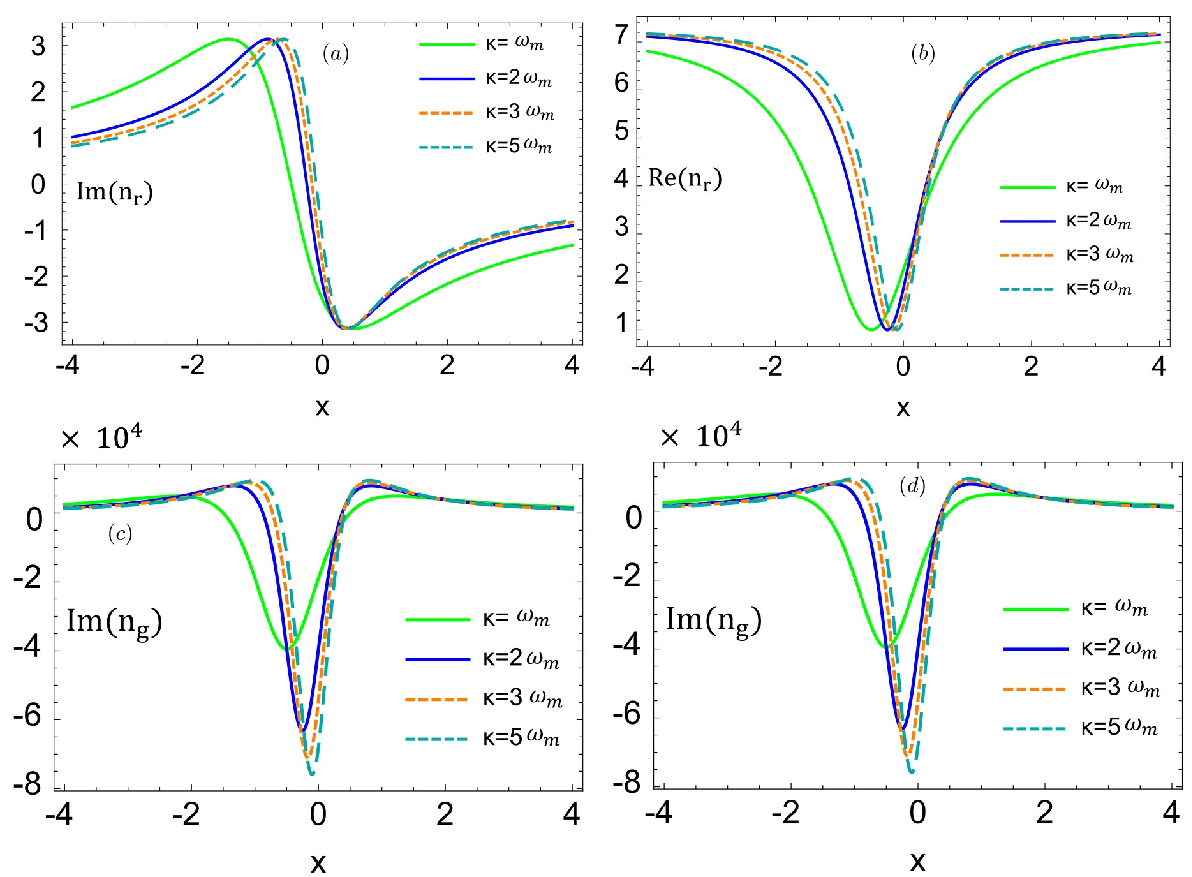}
%    \put(-110,150){($ a $)} \quad
% % \subfloat[The Real of refractive index ]
% { \includegraphics[width=.46\textwidth]{Fig4b.eps}}
% \put(-110,150){($ b $)} \quad \\
%  %\subfloat[The real part of group index index ]
%  { \includegraphics[width=.47\textwidth]{Fig4c.eps}}
%  \put(-200,140){($ c $)} \quad
%  %\subfloat[The imaginary part of the rgroup index ]
%  { \includegraphics[width=.47\textwidth]{Fig4d.eps}}
%   \put(-120,160){($ d $)} \quad
 \caption{(a)(c) The imaginary and (b)(d) the real part of the (a)(b) refractive index and (c)(d) the group index as a function of $x$ for different value of cavity decay rate $\kappa$. Here, we take $\gamma_{m}=10^{-4}\omega_{m}$.}
 \label{fig:2}
\end{figure}
Next, the group index of the optomechanical system versus probe detuning is reported in Fig. 2 (c,d). In Fig. 2(c), the imaginary part of the group index shows normal (positive) and anomalous (negative) dispersion for all damping momentum rates. Therefore, we can see an enhanced anomalous dispersion at $x=0.5$ and normal dispersion around resonance. For low damping rates, see the green curve of Fig. 2(c), we observe a group index of -80000 for $\gamma_{m}=0.5 $  and -20000 for $\gamma_{m}=2$, which is an enhancement of 4 times. This means that the optical properties of the optomechanical system are prominent for small values damping rate of the momentum. Interestingly, for a low damping rate, a dip in the dispersion is produced for a very narrow bandwidth, allowing for a quick switch anomalous to normal around the resonance.
The real group index (absorption and gain profiles) versus probe detuning for various momentum damping rates are reported in Fig. 2(d). We notice the gain of the probe light along the negative probe detuning and absorption along the positive probe detuning (almost for all values of damping momentum rates). The enhanced gain and absorption of the probe light are observed for weak damping rates, see the green curve of Fig. 2(d). Indeed, we can switch from absorption to gain by manipulating the detuning. Both profiles of gain and absorption are symmetric (the sum of the curves is vanishing), leading to a similar parity-time (PT) symmetric behavior in the opto-mechanical system.
\begin{figure}[tbp]
\centering
  %\subfloat[The light drag versus y]
\includegraphics[width=1\columnwidth,height=2.7in]{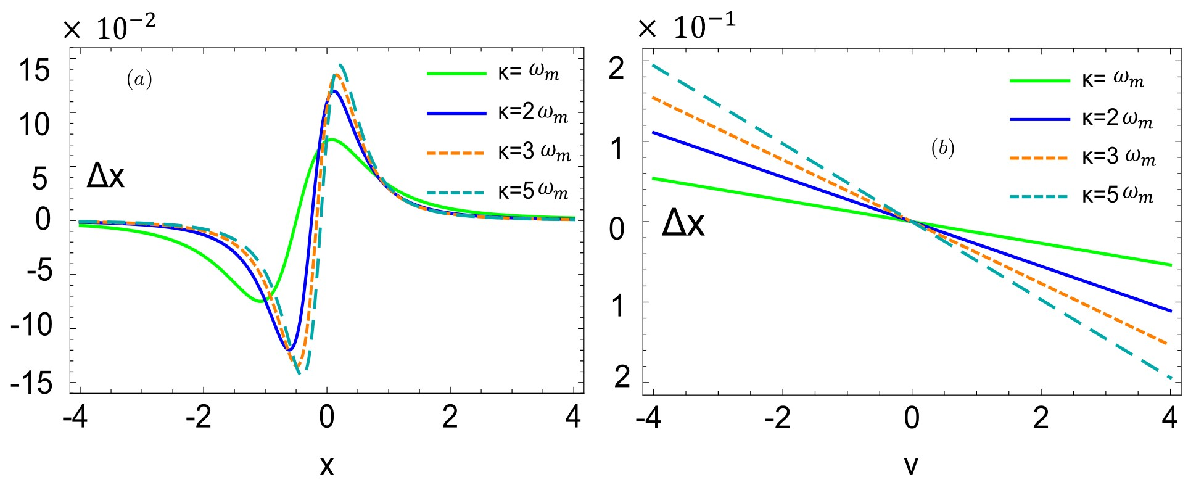}
%\put(-200,170){($ a $)} \quad
%  %\subfloat[The light darg versus v ]
%{\includegraphics[width=.49\textwidth,height=2.8in]{Fig5b.eps}}\put(-110,140){($ b $)} \quad
 \caption{The light drag in an optomechanical system as a function of (a) $x$ and (b) $v$  for different values of cavity decay rate $\kappa$. The rest of the parameters are the same as in Fig. 4.}\label{fig:6}
\end{figure}
\begin{figure}[b!]
\centering
  %\subfloat[The imaginary of refractive index versus probe detuning]
  \includegraphics[width=1\columnwidth,height=5in]{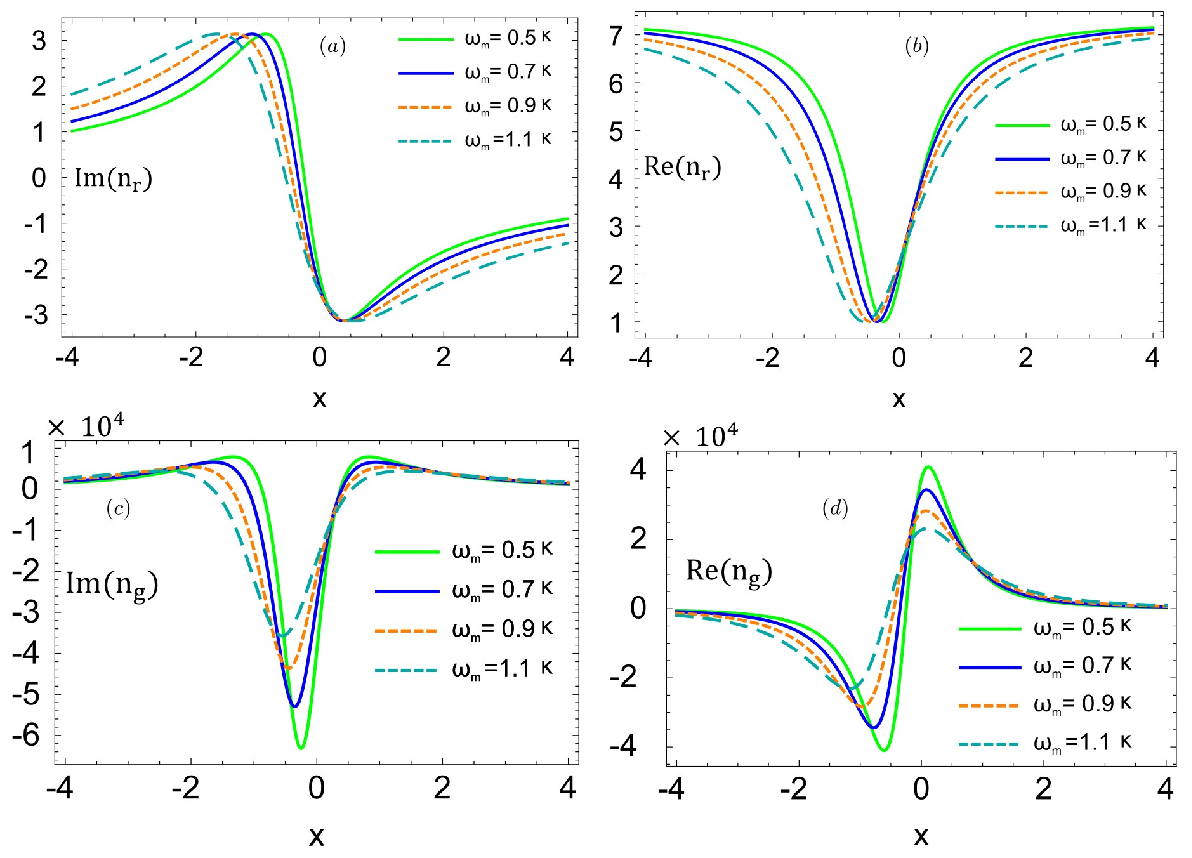}
  %  \put(-110,150){($ a $)} \quad
% % \subfloat[The Real of refractive index ]
% { \includegraphics[width=.46\textwidth]{Fig6b.eps}}
% \put(-110,150){($ b $)} \quad \\
%  %\subfloat[The real part of group index index ]
%  { \includegraphics[width=.47\textwidth]{Fig6c.eps}}
%  \put(-200,140){($ c $)} \quad
%  %\subfloat[The imaginary part of the rgroup index ]
%  { \includegraphics[width=.47\textwidth]{Fig6d.eps}}
%   \put(-150,140){($ d $)} \quad
 \caption{(a)(c)The imaginary and (b)(d) the real part of the (a)(b) refractive index and (c)(d) the group index as a function of $x$ for different value of mechanical frequency $\omega_{m}$. The rest of the parameters are the same as in Fig. 4.}\label{fig:2}
\end{figure}

In Fig. 3(a), we obtain negative drag for negative detuning and positive drag for positive detuning, where a low momentum damping rate leads to an enhanced light drag around the resonance. Besides, for $\gamma =1$, light drag is 0.04mm, and it is increased twice for $\gamma =0.5$. In contrast, in Fig. 3(b), for $\gamma=0.5$, we have positive light drag, and it is in the opposite direction to the medium velocity. Additionally, it has a negative slope, suggesting a superluminal behavior of the probe light in the system.
Finally, it is worth noting that the momentum damping rate can be experimentally controlled by detuning the laser frequency and amplitude, which increases radiation pressure on the mirror and contributes to damping rate control via optomechanical interactions \cite{cripe2020radiation}. Furthermore, the desired damping rate can be controlled by the position of the mirror using techniques such as force sensing with feedback loops \cite{bemani2022force}.

Next, in Fig. 4, we report the dependence of the refractive and group indices on $\kappa$, where $\kappa$ is the decay of the cavity. For small cavity length, we have a faster cavity decay i,e $\kappa \propto \frac{1}{L}$ (the time for a photon round trip in the cavity is linked to its length). We see in Fig. 4(a) and 4(b) that a small cavity decay $\kappa$ leads to a shift in the dispersion and absorption towards the negative detuning around the resonance. On the other side, in Fig. 4(c) and 4(d), we observe the normal and anomalous dispersion (related to the slope of the group velocity). Hence, for a high cavity decay rate $\kappa$, both the anomalous and normal dispersion get steeper around the resonance, and the dip reaches -80000. In Fig. 4(d), it is clear that the gain and absorption depend on the cavity decay rate. In general, the cavity decay is inversely proportional to the cavity photon lifetime ($\tau$). Therefore, absorption and gain mechanisms are affected by the light-matter interaction through photon emissions, saturation, optical cooling, losses, and excess heating. In this context, fine-tuning the cavity's decay characteristics can be experimentally engineered by adjusting the cavity's length, applying a mirror coating to affect its reflectivity, using optical attenuators, or adding tunable losses to the cavity through the use of electro-optic or acousto-optic modulators.

In Fig. 5, the light shift is sensitive to the cavity decay rates $\kappa$ (see the blue, red, and dashed-cyan curves). The light drag reaches about -0.15 and vanishes out of the resonance. Additionally, for negative velocity, we have a positive shift indicating superluminal behavior of the light (and vice versa for positive detuning) [see Fig. (b)]. The slope of the curves indicates the rate of variation of the light drag due to the change in velocity of the medium. Thus, we observe higher slopes for higher cavity decay rates (for instance, consistent with cavity length variations).
\begin{figure}
\centering
  %\subfloat[The light drag versus y]
\includegraphics[width=1\columnwidth,height=2.7in]{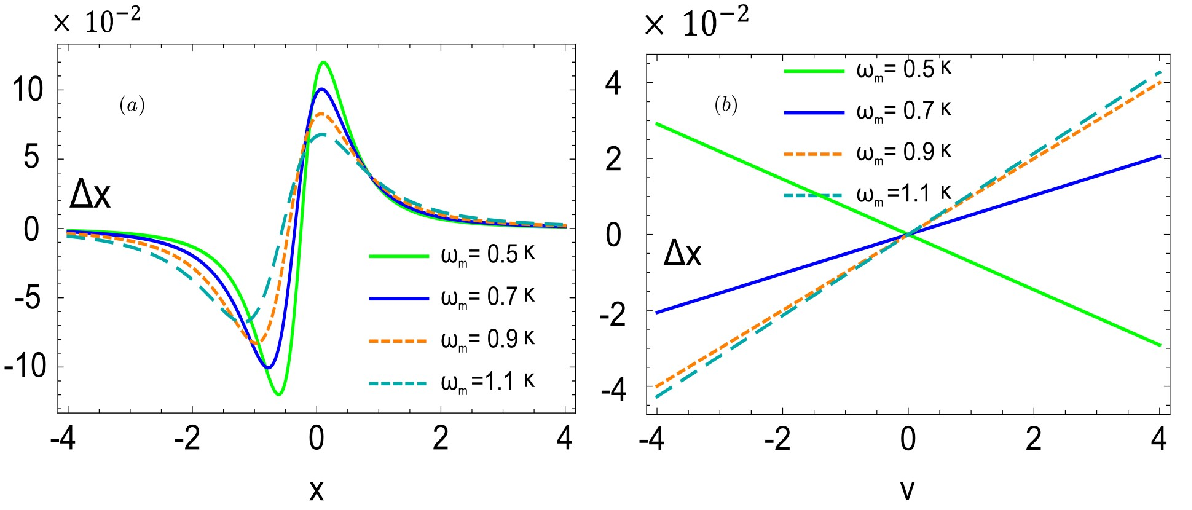}
 %\put(-200,170){($ a $)} \quad
%  %\subfloat[The light darg versus v ]
%{\includegraphics[width=.49\textwidth,height=3in]{Fig7b.eps}}\put(-200,170){($ b $)} \quad
 \caption{The light drag in an optomechanical system as a function of (a) $x$ and (b) $v$  for different values of mechanical frequency $\omega_{m}$. The rest of the parameters are the same as in Fig. 4.}
\label{fig:6}
\end{figure}
Figure 6 reports the refractive and the group indexes versus the probe detuning under the effect of the optomechanical oscillations. The optomechanical cavity often requires strong coupling between the optical field and the cavity; hence, $\omega_{m}$ has to be high enough to avoid excessive thermal noise, which can dominate at lower frequencies. Clearly, $\omega_{m}$ and $\gamma_{m}$ play inverse roles for light drag in the cavity; hence, dispersion is lower for high $\gamma_{m}$ and shifts towards negative detuning whereas it is higher for high $\omega_{m}$ and gets closer to the resonance (likewise for the absorption spectra). Controlling the properties of the system under the effect of $\omega_{m}$ is crucial for applications involving precision sensing and quantum information processing, among others. Thus, in Fig. 6, it is clear that the mirror's frequency affects the dispersion shift towards the negative detuning, and we obtain a steeper curve for the low value of the mechanical frequency. In contrast, an inverse behavior (in comparison with the decay rate effect ) is observed in Fig. 6(c) and Fig. 6(d). For instance, higher frequencies enable lower depth in the dispersion profile. Likewise, the gain and absorption are prominent for lower frequencies [see Fig. 6(d)].
The lateral light drags in terms of detuning $\Delta_{x}$ is reported in Fig. 7. (a) for various $\omega_{m}$ rates, i.e., the mirror mechanical frequency. Additionally,  the light-drag is analyzed in terms of the transnational velocity, as depicted in Fig. 7.(b). Positive and negative lateral shifts are observed for positive and negative detuning, respectively see Fig. 7(a). Moreover, low shifts in light drag occur for higher mechanical frequency, i.e., 0.06 m for $\omega_{m}= 11000$ Hz. The light drags gradually increase with a gradual decrease of the mechanical frequency, and we obtain almost twice the light drag, i.e., 0.012 m when the mechanical frequency is reduced to  $\omega_{m}= 5000$ Hz. In Fig. 7(b), curves representing lateral drag exhibit a range of slopes, from positive to negative, that change from negative to positive. Subluminal light transmission through the cavity is demonstrated when we increase the mirror frequency to 7000 Hz.
\begin{figure}[tbp]
\centering
  %\subfloat[The light drag versus y]
 \includegraphics[width=1\columnwidth,height=2.7in]{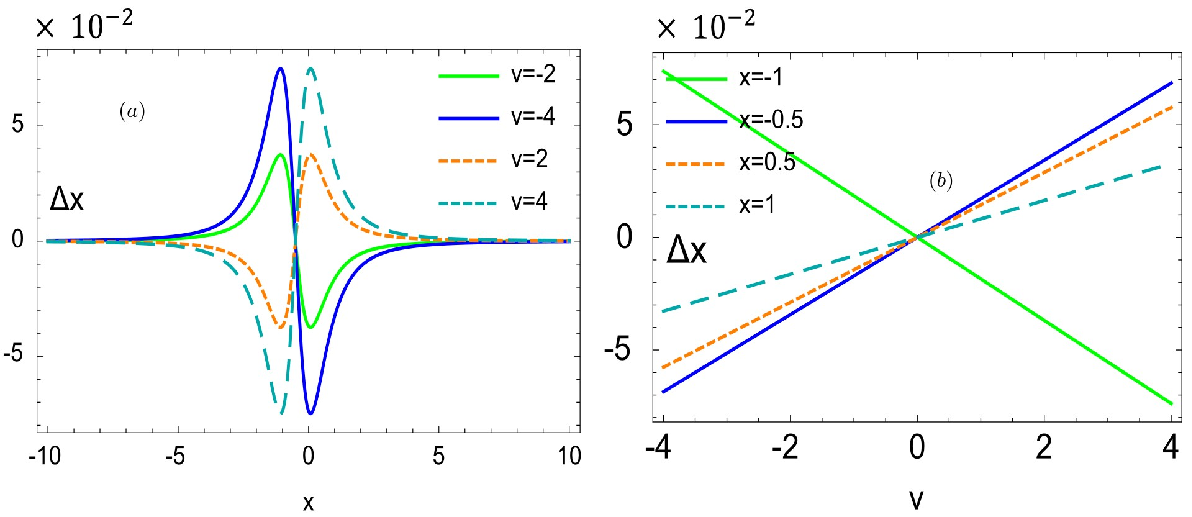}
   %\put(-200,170){($ a $)} \quad
%  %\subfloat[The light darg versus v ]
%{\includegraphics[width=.49\textwidth,height=3in]{Fig8b.eps}}\put(-110,140){($ b $)} \quad
 \caption{The light drag in an optomechanical system as a function of (a) $x$ and (b) $v$  for different values of (a) velocity (b) $x$.}\label{fig:8}
\end{figure}

By varying the transnational velocity's amplitude and direction, we examine the relationship between lateral light drag and probe detuning [see Fig. 8(a)]. The red curve in Fig. 8(a) shows that at a velocity of 2m/s, there is a positive light drag of 0.03m at $\Delta_{p}= 0.3 \gamma$ and a negative light drag of 0.03m at $\Delta_{p}= -\gamma$. We notice nearly twice as much positive and negative light drag (see cyan curve) when the velocity is increased to 4m/s. When the medium flows in the other direction, we see exactly the opposite behavior of the light drag, see the green and blue curve in Fig. 8(a). Figure 8(b) shows the light drag versus transnational velocity in terms of detuning. For negative probe detuning, we see a negative slope, and for positive detuning, a positive slope. Therefore, we can change the light's propagation through the optomechanical cavity from super- to sub-luminal and from sub- to super-luminal by adjusting the probe's detuning. We recall that the change in the transnational velocity of the mirror causes a Doppler shift in the frequency of the light \cite{wayne2013relationship} (the mirror motion toward/away from the light affects the reflected light frequency). Thus, when the resonance requirement is no longer precisely met, the cavity may become detuned as a result of this Doppler shift. This detuning modifies the interaction between the light and the system, which also may result in a feedback loop \cite{altmann2019alignment} that further affects the motion of the mirror and further influences the mirror's motion.
In addition, the optomechanical interaction may experience nonlinear effects \cite{xu2015quantum}, if the velocity change is high. The light drag effect can be further influenced by complex dynamical behavior resulting from these nonlinearities, such as bistability \cite{zhang2017bistable}, or mode splitting \cite{zhao2023controllable}.

\section{Conclusions}\label{sec4}
This work demonstrated that a mechanical resonator coupled to an optical cavity can experience light drag. Thus, we analyzed the refractive and group indexes, both related to light drag. The damping rate/frequency of the mechanical resonator, and the cavity loss enabled complete quantum control over the light shift in the system. Our observations revealed that applying a phase shift in the detuning allows up to 90$\%$ transmission of the probe light. We also observed enhanced anomalous dispersion near resonance, with light drag influenced by the cavity decay rate. Finally, by varying the amplitude and direction of the translational velocity, we found that lateral light drag depends on detuning, enabling us to switch the light's propagation through the optomechanical cavity from superluminal to subluminal and vice versa by adjusting the probe's detuning. Our demonstration of light drag in an optomechanical system not only paves the way for the next generation of photonic devices but also suggests enhancing the sensitivity of optomechanical systems. Therefore, we claim that our technique has the potential to be utilized with present-day technology in quantum information processing.
\section*{APPENDIX A: DERIVATION OF}
To solve Eq. (2), we substitute the formal solution
$\left\langle s\right\rangle =s_{0}+\varepsilon _{p}e^{-i\delta t}s_{+}+\varepsilon
_{p}^{\ast }e^{i\delta t}s_{-}(s=q,p,c)$
\begin{equation}
q_{\circ }=\frac{\chi _{\circ }\left\vert c_{\circ }\right\vert ^{2}}{%
m\omega _{m}^{2}},
\end{equation}
\begin{equation}
q_{+}=\frac{\chi _{\circ }\left( c_{\circ }c_{-}^{\ast }+c_{\circ }^{\ast
}c_{+}\right) }{m\left( \omega _{m}^{2}-i\delta \gamma -\delta ^{2}\right) } \label{qp},
\end{equation}
\begin{equation}
q_{-}=\frac{\chi _{\circ }\left( c_{\circ }c_{+}^{\ast }+c_{\circ }^{\ast
}c_{-}\right) }{m\left( \omega _{m}^{2}+i\delta \gamma -\delta ^{2}\right) }. \label{qn}
\end{equation}

\begin{equation}
c_{\circ }=\frac{\varepsilon _{c}}{\kappa +i\Delta },
\end{equation}

\begin{equation}
c_{+}=\frac{ic_{\circ }q+\frac{\chi _{\circ }}{\hbar }+1}{\kappa +i\left(
\Delta -\delta \right) }, \label{cp}
\end{equation}

\begin{equation}
c_{-}=\frac{ic_{\circ }q-\frac{\chi _{\circ }}{\hbar }}{\kappa +i\left(
\Delta +\delta \right) }. \label{cn}
\end{equation}
It can be seen from Eq. (\ref{qp}) and Eq. (\ref{qn}), $q_{+}=q_{-}^{*}$. Furthermore, using Eq. (\ref{qp}) and Eq. (\ref{cn}), one can obtained
\begin{equation}
c_{\circ }c_{-}^{\ast }=\frac{M}{1-M}c_{\circ }^{\ast }c_{+},\label{CC}
\end{equation}
where
\begin{equation}
M=\frac{-i\left\vert c_{\circ }\right\vert ^{2}\chi _{\circ }^{2}}{m\hbar
\left( \omega _{m}^{2}-i\delta \gamma -\delta ^{2}\right) \left( \kappa
-i\left( \Delta +\delta \right) \right) }.
\end{equation}
Simultaneously solving Eq. (\ref{qp}), Eq. (\ref{cp}) and Eq. (\ref{CC}), we obtain the following expression of $c_{+}$
\begin{equation}
c_{+}=\frac{1}{\kappa -i\left( \Delta +\delta \right) +\frac{\beta }{\frac{%
\left( \delta ^{2}-\omega _{m}^{2}+i\delta \gamma \right) }{2i\omega _{m}}-%
\frac{\beta }{\kappa -i\left( \Delta +\delta \right) }}}. \label{FE}
\end{equation}
Near the resonance condition, i.e. $\Delta \sim\delta \sim\omega _{m}$, we have $\delta +\Delta \sim 2\omega _{m}$ and $\delta ^{2}-\omega _{m}^{2}\sim 2\omega _{m}\left( \delta -\omega
_{m}\right)$. Furthermore, if we set $x= \delta -\omega _{m}$, simplifying Eq. (\ref{FE}) yields
\begin{equation}
c_{+}=\frac{1}{\kappa -ix+\frac{\beta }{\frac{\gamma }{2}-ix+\frac{\beta }{%
\kappa -2i\omega _{m}}}}.
\end{equation}
Finally, by multiplying the above expression of $c_{+}$ by $2\kappa$, we obtain the absorption and dispersion quadrature of the output field $\varepsilon _{T}$.

\section*{Acknowledgement}
Amjad Sohail acknowledges the financial support given by CNPq through grant number 171707/2023-0.
\section*{Data availability}
All data supporting the findings in this study is available in the article.
%\section*{Author contributions statement}
%A.S and H.A. conceptualization, N.B methodology. A.S analysis. All authors contributed to discussions and writting. All authors reviewed the manuscript.
\section*{ORCID iD}
Hazrat Ali https://orcid.org/0000-0003-1957-3629\\
Nadia Boutabba https://orcid.org/0000-0001-8867-5464 \\
Amjad Sohail https://orcid.org/0000-0001-8777-7928

%\bibliography{optomech1}

\begin{thebibliography}{99}
\bibitem{fresnel1818influence}Fresnel A J, Sur \'{I}influence du mouvement terrestre dans quelques phenomenes d'optique: lettre de M. Fresnel a M. Arago. 1818 \textit{Ann. Chim. Phys.} \textbf{9} 57–66.
\bibitem{banerjee2022anomalous} Banerjee C, Solomons Y, Black A N, Marcucci G, Eger D, Davidson N, Firstenberg O and Boyd R W, Anomalous optical drag 2022 \textit{Phys. Rev.
research} \textbf{4} 033124.
\bibitem{qin2020fast}Qin T, Yang J, Zhang F, Chen Y, Shen D, Liu W, Chen L, Jiang X, Chen X and Wan W, Fast-and slow-light-enhanced light drag in a moving microcavity 2020 \textit{Commun. Phys.} \textbf{3} 118.
\bibitem{ullah2023coherent} Ullah I, Munir S, Khan I, Khan A, Bacha B A and Ahmad W, Coherent manipulation of anomalous optical drag in atomic medium 2023 \textit{Opt. and Quantum Electronics} \textbf{55} 757.
\bibitem{fizeau1991hypotheses} Fizeau M, Sur les hypoth\`{e}ses relatives à \`{I}\'{e}ther lumineux, et sur une exp\'{e}rience qui parait d\'{e}montrer que le mouvement des corps change la vitesse avec laquelle la lumi\`{e}re se propage dans leur int\'{e}rieur 1991 \textit{SPIE milestone series} \textbf{28} 445–449.
\bibitem{zeeman1919propagation} Zeeman P, The propagation of light in moving transparent solid substances. I. Apparatus for the observation of the fizeau-effect in solid substance 1919 KNAW, \textit{Proc.} \textbf{22} 1919–1920.
\bibitem{kox1993pieter} Kox A, Pieter Zeeman's Experiments on the Equality of Inertial and Gravitational Mass 1993 \textit{Einstein Studies} \textbf{5} 173–181.
\bibitem{safari2016light} Safari A, \textit{et. al.}, Light-drag enhancement by a highly dispersive rubidium vapor 2016 \textit{Phys. Rev. lett.} \textbf{116} 013601.
\bibitem{ali2023light} Ali H and Boutabba N 2023 \textit{Phys. Scr.} \textbf{98} 045410.
\bibitem{boutabba2023light} Boutabba N, Rasheed Z and Ali H, Light drag in a left-handed atomic medium via Cross Kerr-like nonlinearity 2023 \textit{Chaos, Solitons and Fractals} \textbf{177} 114165.
\bibitem{kuan2016large} Kuan P C, Huang C, Chan W S, Kosen S and Lan S Y, Large Fizeau's light-dragging effect in a moving electromagnetically induced transparent medium 2016 \textit{Nat. Commun.} \textbf{7} 13030.
\bibitem{o2022random} \'{O}Sullivan J, Kennedy O W, \textit{et al.} Random-access quantum memory using chirped pulse phase encoding 2022 \textit{Phys. Rev. X} \textbf{12} 041014.
\bibitem{radeonychev2020observation} Radeonychev Y, Khairulin I, Vagizov F, Scully M and Kocharovskaya O, Observation of Acoustically Induced Transparency for $\gamma$-Ray Photons 2020 \textit{Phys. Rev. Lett.} \textbf{124} 163602.
\bibitem{scully2003playing} Scully M O and Zubairy M S, Playing tricks with slow light 2003 \textit{Science} \textbf{301} 181–182.
\bibitem{sete2012controllable} Sete E A and Eleuch H, Controllable nonlinear effects in an optomechanical resonator containing a quantum well 2012 \textit{Phys. Rev. A} \textbf{85} 043824.
\bibitem{agarwal2010electromagnetically} Agarwal G S and Huang S, Electromagnetically induced transparency in mechanical effects of light 2010 \textit{Phys. Rev. A—Atomic, Molecular, and Optical Physics} \textbf{81} 041803.
\bibitem{huang2011electromagnetically} Huang S and Agarwal G, Electromagnetically induced transparency from two-phonon processes in quadratically coupled membranes 2011 \textit{Phys. Rev. A—Atomic, Molecular, and Optical Physics} \textbf{83} 023823.
%\bibitem{sohail2016optomechanically} Sohail A, Zhang Y, Zhang J and Yu C s 2016 \textit{Sci. Rep.} \textbf{6} 28830.
\bibitem{yan2020optomechanically} Yan X B, Optomechanically induced transparency and gain 2020 \textit{Phys. Rev. A} \textbf{101} 043820.
\bibitem{xiong2018fundamentals} Xiong H and Wu Y, Fundamentals and applications of optomechanically induced transparency 2018 \textit{Appl. Phys. Rev.} \textbf{5} 031305.
\bibitem{kristensen2024long} Kristensen M B, Kralj N, Langman E C and Schliesser A, Long-lived and Efficient Optomechanical Memory for Light 2024 \textit{Phys. Rev. Lett.} \textbf{132} 100802.
\bibitem{weis2010optomechanically} Weis S, Rivi\`{e}re R, Del\'{e}glise S, Gavartin E, Arcizet O, Schliesser A and Kippenberg T J, Optomechanically induced transparency 2010 \textit{Science} \textbf{330} 1520–1523.
\bibitem{yang2019optomechanically} Yang X, Yin Z and Xiao M, Optomechanically induced entanglement 2019 \textit{Phys. Rev. A} \textbf{99} 013811.
\bibitem{koppenhofer2023single} Koppenh\"{o}fer M, \textit{et. al.}, Single-spin readout and quantum sensing using optomechanically induced transparency 2023 \textit{Phys. Rev. Lett.} \textbf{130} 093603.

\bibitem{kocharovskaya2001stopping} Kocharovskaya O, Rostovtsev Y and Scully M O, Stopping light via hot atoms 2001 \textit{Phys. Rev. Lett.} \textbf{86} 628.
\bibitem{turukhin2001observation} Turukhin A, Sudarshanam V, Shahriar M, Musser J, Ham B and Hemmer P, Observation of ultraslow and stored light pulses in a solid 2001 \textit{Phys. Rev. Lett.} \textbf{88} 023602.
\bibitem{sohail2023rotational} Amjad S, Arif R, Akhtar N, Ziauddin, Peng J X, Xianlong G and Gu Z, A rotational-cavity optomechanical system with two revolving cavity mirrors: optical response and fast-slow light mechanism 2023 \textit{Eur. Phys. J. Plus} \textbf{138} 417.
%\bibitem{sohail2020enhancement} Sohail A, Rana M, Ikram S, Munir T, Hussain T, Ahmed R and Yu C s 2020 \textit{Quantum Inf. Process.} \textbf{19} 1–18.
\bibitem{ahmed2017optomechanical} Ahmed R and Qamar S, Optomechanical entanglement via non-degenerate parametric interactions 2017 \textit{Phys. Scr.} \textbf{92} 105101.
\bibitem{ahmed2019effects} Ahmed R and Qamar S, Effects of laser phase noise on optomechanical entanglement in the presence of a nonlinear Kerr downconverter 2019 \textit{Phys. Scr.} \textbf{94} 085102.
\bibitem{sohail2023enhanced23} Sohail A, Abbas Z, Ahmed R, Shahzad A, Akhtar N and Peng J X, Enhanced entanglement and controlling quantum steering in a Laguerre–Gaussian cavity optomechanical system with two rotating mirrors 2023 \textit{Ann. der Physik} \textbf{535} 2300087.
\bibitem{he2015sensitivity} He Y, Sensitivity of optical mass sensor enhanced by optomechanical coupling 2015 \textit{Appl. Phys. Lett.} \textbf{106} 121905.
\bibitem{lobino2009memory} Lobino M, Kupchak C, Figueroa E and Lvovsky A, Memory for light as a quantum process 2009 \textit{Phys. Rev. Lett.} \textbf{102} 203601.
\bibitem{yan2021optomechanically2} Yan X B, Optomechanically induced ultraslow and ultrafast light 2021 \textit{Physica E: Low-dimensional Systems and Nanostructures} \textbf{131} 114759.
\bibitem{safavi2011electromagnetically} Safavi-Naeini A H, \textit{et. al.}, Electromagnetically induced transparency and slow light with optomechanics 2011 \textit{Nature} \textbf{472} 69–73.
\bibitem{liu2019nonreciprocal} Liu J H, Yu Y F and Zhang Z M, Nonreciprocal transmission and fast-slow light effects in a cavity optomechanical system 2019 \textit{Opt. Express} \textbf{27} 15382–15390.
\bibitem{li2016transparency} Li L, Nie W and Chen A, Transparency and tunable slow and fast light in a nonlinear optomechanical cavity 2016 \textit{Sci. Rep.} \textbf{6} 35090.

%\bibitem{boutabba2013slowing} Boutabba N and Eleuch H 2013 \textit{Applied Mathematics \& Information Sciences} \textbf{7} 1505.
    \bibitem{akram2015tunable} Akram M J, Khan M M and Saif F, Tunable fast and slow light in a hybrid optomechanical system 2015 \textit{Phys. Rev. A} \textbf{92} 023846.

\bibitem{WLM} Liu W, Abbas M, Asadpour S H, Hamedi H R, Zhang P and Barry C Sanders, Generating grating in cavity magnomechanics 2024 \textit{New J. Phys.} \textbf{26} 093042.

\bibitem{jiang2017fano} Jiang C, Jiang L, Yu H, Cui Y, Li X and Chen G, Fano resonance and slow light in hybrid optomechanics mediated by a two-level system 2017 \textit{Phys. Rev. A} \textbf{96} 053821.
%\bibitem{sohail2017controllable} Sohail A, Zhang Y, Usman M and Yu C s 2017 \textit{EPJD} \textbf{71} 1–6.
\bibitem{baraillon2020linear} Baraillon J, Taurel B, Labeye P and Duraffourg L, Linear analytical approach to dispersive, external dissipative, and intrinsic dissipative couplings in optomechanical systems 2020 \textit{Phys. Rev. A} \textbf{102} 033509.
\bibitem{cripe2020radiation} Cripe J and Cripe J 2020 Broadband Measurement and Reduction of Quantum Radiation Pressure Noise in the Audio Band 67–79.
\bibitem{bemani2022force} Bemani F, Cernot\'{\i}k O, Ruppert L, Vitali D and Filip R, Force sensing in an optomechanical system with feedback-controlled in-loop light 2022 \textit{Phys. Rev. Applied} \textbf{17} 034020.

\bibitem{wayne2013relationship} Wayne R, The relationship between the optomechanical Doppler force and the magnetic vector potential 2013 \textit{The African Review of Physics} \textbf{8} 0042.
\bibitem{altmann2019alignment} Altmann B, Betker T, Pape C and Reithmeier E, Alignment strategy for an optomechanical image derotator using a laser Doppler vibrometer 2019 \textit{Applied optics} \textbf{58} 6555–6568.
\bibitem{xu2015quantum} Xu X, Gullans M and Taylor J M, Quantum nonlinear optics near optomechanical instabilities 2015 \textit{Phys. Rev. A} \textbf{91} 013818.
\bibitem{zhang2017bistable} Zhang W Z, Li W L, Cheng J and Mu Q, Bistable cooling in optomechanical system 2017 \textit{arXiv preprint} arXiv:1710.11308.
\bibitem{zhao2023controllable} Zhao G, Zhu J, Hou J, Chen Y, Lin J, Cheng Y, Chen X, Zheng Y and Wan W, Controllable EIT-like mode splitting in a chiral microcavity 2023 \textit{Opt. Lett.} \textbf{48} 755–758.
\end{thebibliography}
%\bibliographystyle{iopart-num}
\end{document}